\documentclass[prb, twocolumn, preprintnumbers ,amsmath, amssymb, superscriptaddress, aps, floatfix, longbibliography]{revtex4-2}

\usepackage{color}
\usepackage[T1]{fontenc}
\usepackage{amsmath,amssymb}
\usepackage{pifont}
\usepackage{amssymb}  
\usepackage{bbold}
\usepackage{float}
\usepackage{tikz}
\usepackage{caption}
\usepackage{subcaption}
\usepackage{makecell}
\usepackage{pifont}   
\usepackage{graphicx} 
\usepackage{dcolumn}  
\usepackage{bm}       
\usepackage{multirow} 
\usepackage[colorlinks=true, linkcolor=blue, citecolor=blue]{hyperref}
\usepackage{placeins}

\usepackage{graphicx}
\graphicspath{{figs/}}

\begin{document}
\title{Deep learning of deformation-dependent conductance in thin films: nanobubbles in graphene}
\author{Jack G. Nedell}
\affiliation{Department of Physics, Northeastern University, Boston, MA 02115, USA}
\author{Jonah Spector}
\affiliation{Department of Physics, Northeastern University, Boston, MA 02115, USA}
\author{Adel Abbout}
\affiliation{Department of Physics, King Fahd University of Petroleum and Minerals, 31261 Dhahran, Saudi Arabia}
\author{Michael Vogl}
\affiliation{Department of Physics, King Fahd University of Petroleum and Minerals, 31261 Dhahran, Saudi Arabia}
\author{Gregory A. Fiete}
\affiliation{Department of Physics, Northeastern University, Boston, MA 02115, USA}
\affiliation{Department of Physics, Massachusetts Institute of Technology, Cambridge, MA 02139, USA}
\date{\today}

\begin{abstract}
  Motivated by the ever-improving performance of deep learning techniques, we design a mixed input convolutional neural network approach to predict transport properties in deformed nanoscale materials using a height map of deformations, as can be obtained from scanning probe measurements, as input.  We employ our approach to study electrical transport in a graphene nanoribbon deformed by a number of randomly positioned nano-bubbles.  Our network is able to make conductance predictions valid to an average error of 4.3\%. We find that such low average errors are achieved by a redundant input of energy values, yielding predictions that are 30-40\% more accurate than conventional architectures. We demonstrate that the same method can learn to predict the valley-resolved conductance, with success specifically in identifying the energy at which inter-valley scattering becomes prominent. We demonstrate the robustness of the approach by testing the pre-trained network on samples with deformations differing in number and shape from the training data. We furthermore employ a graph theoretical analysis of the structure and outputs of the network and conclude that a tight-binding Hamiltonian can be effectively encoded in the first layer of the network, which is supported by numerical findings. Our approach contributes a new theoretical understanding and a refined methodology to the application of deep learning for the determination of transport properties based on real-space disorder information.

\end{abstract}
\maketitle
\section{Introduction}

Massive progress in the accuracy and computational efficiency of deep learning techniques, combined with widespread application of these methods, has rendered deep learning an increasingly viable tool for complex problems in physics \cite{tanaka_deep_2021,mehta_high-bias_2019,10.21468/SciPostPhysLectNotes.29,Carleo_2019}. This can be seen in the numerous recent applications of deep learning; a prolific and successful example has been in data analysis at the LHC \cite{abdughani_probing_2019,verma_jet_2021}, and applications in condensed matter and adjacent fields have prospered, too \cite{beer_training_2020,sun_deep_2018,PhysRevX.8.031084,Bedolla_2020,Carrasquilla_2020,Schmidt2019}. A particular topic of interest has been the prediction and identification of phase transitions
 \cite{carrasquilla_machine_2017,schoenholz_structural_2016,rem_identifying_2019, suchsland_parameter_2018,PhysRevLett.120.257204,PhysRevB.103.134203}. Among the most common deep learning techniques, also employed in this work, is the convolutional neural network (CNN), which is also the standard class of neural networks used for image recognition. CNNs are favored for their versatility, and the implementation of 2D or 3D convolutions allows these networks to map multidimensional data to almost any correlated quantity.

  \begin{figure}[b]
    \centering
    \includegraphics[width=.5\textwidth]{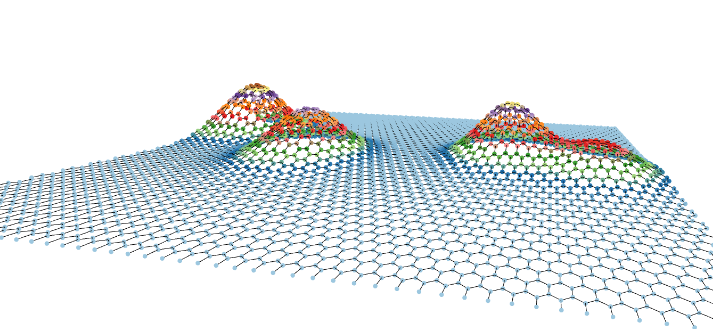}

    \caption{ 
    Illustrated example of the type of deformed graphene nanoribbon systems we consider in this work.}
    
    
    \label{fig:sys}
\end{figure}

Here, we show that a fast, accurate prediction of the transport properties of deformed graphene can be obtained from only a height map by applying a mixed input neural network that includes a CNN branch. An illustration of an example deformed graphene system is included in Fig. \ref{fig:sys}.  A deformation height map can be obtained in an experimental setting with standard imaging techniques, such as scanning tunneling microscopy, making this methodology feasible for an industrial application.
 \begin{figure*}[t]
    \centering
    \includegraphics[width=.8\textwidth]{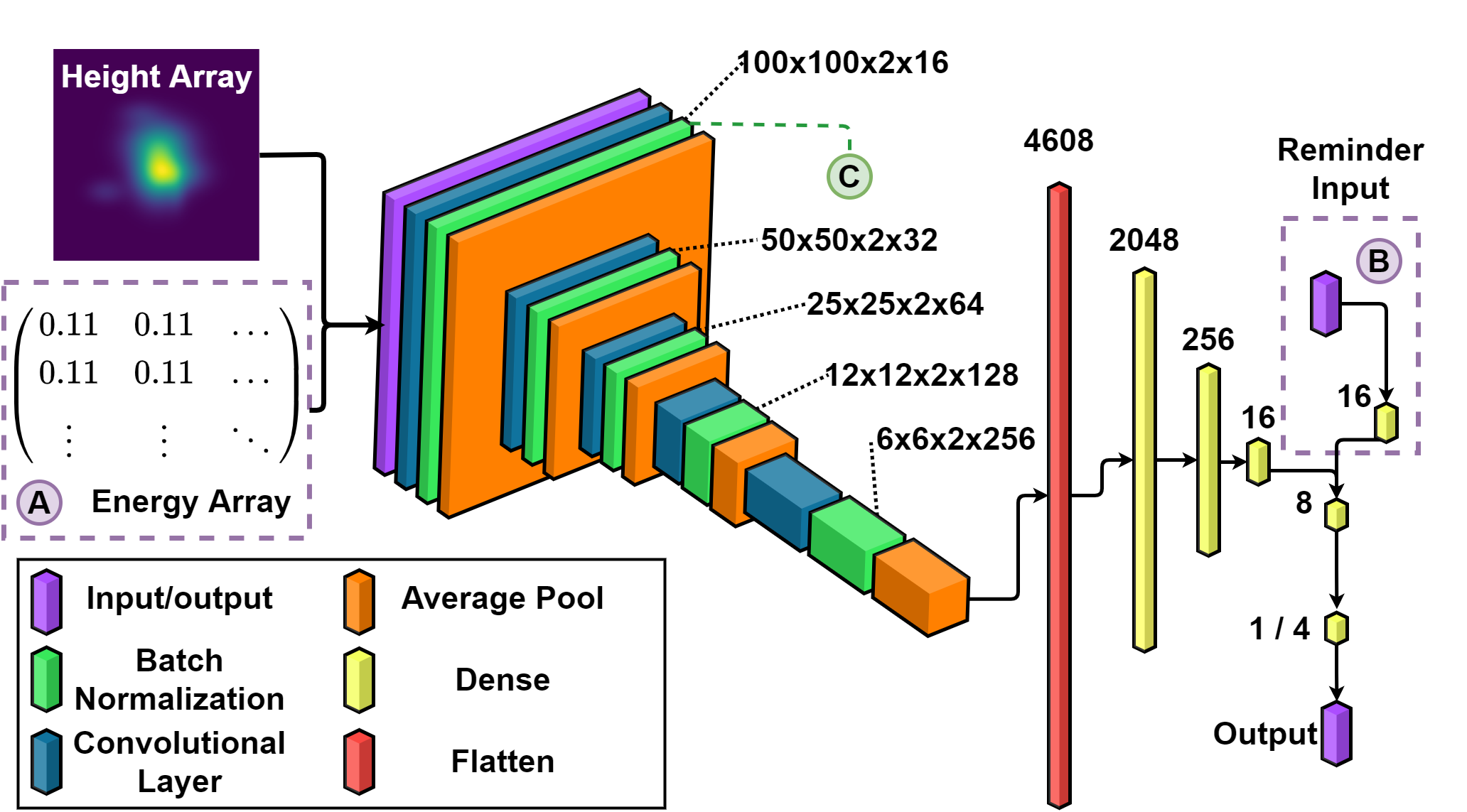}

    \caption{ 
    Diagram showing the architecture of the neural network. The convolutional kernel size was (3,3,2), while the pool size was (2,2,1). Labels A and B correspond to each redundant energy input (See \ref{ssec:redundant}), while C indicates the layer extracted for linear mapping (See \ref{sec:graph_numeric}). In the text we detail results for different variations of these inputs.}
    
    
    \label{fig:net}
\end{figure*}
  Specifically, we will focus our attention on nanoscopic deformations in 2D materials, referred to as nanobubbles. The impact of these nanobubbles on electronic transport can be studied statistically \cite{abbout_statistical_2016} or with standard numeric methods. However, exact analytic treatment of deformations is difficult, spurring work in applying approximations to describe electronic transport in deformed materials such as graphene \cite{stegmann_current_2016,Rost_2019}. One novel physical effect of nano-deformations in graphene is the production of a strong effective magnetic field, up to hundreds of Tesla, which are sensitive to the graphene valley degrees of freedom. Depending on their shapes, these deformations can filter or split the two valleys selectively \cite{PhysRevLett.117.276801}, opening the door to the field of valleytronics \cite{PhysRevLett.110.046601}. Deformed graphene is also of particular interest because it has shown promise for use in numerous applications, such as the ultrasensitive detection of nucleic acids, or as a valley and spin filter \cite{wang_tunable-deformed_2019,hwang_ultrasensitive_2020,settnes_graphene_2016}. Applications such as these, especially if developed at an industrial scale, require reliable and efficient tools--tools faster and less expensive than direct measurements or full calculations--to characterize the physical properties of individual devices. 
  
    In this work, we show that our approach is successful, with a relative error of less than 5\%. With minimal changes our neural network structure is able to make predictions about both the total and valley-resolved components of the conductance, successfully predicting inter-valley scattering. Our neural network architecture was carefully optimized for this class of problems, providing a useful methodology for similar work going forward. We also show that the network is robust against changes to the deformation shape and number of deformations, which is important for non-idealized real world applications.
  
  We additionally analyze the inner-workings of our network to better understand why its predictions are highly accurate. We numerically demonstrate that there exists a simple linear mapping from the first layer of our network to the tight-binding Hamiltonian matrix of the graphene system. The connection between the neural network structure and a tight binding Hamiltonian is studied further on purely theoretical grounds by applying a graph theory analysis. Specifically, we prove the existence of graph morphisms between a convolutional layer in a CNN and a tight-binding Hamiltonian. This fundamental equivalence relation has to our knowledge not been previously identified, and brings valuable insight to the form and function of neural networks and strongly emphasizes the general applicability of our approach.
  
  Before we proceed to discussing the structure of this work let us first take a step back and mention some recent closely related work and how our work differs from it.  Recently there has been work by Peano {\it et al.} \cite{PhysRevX.11.021052} that explored via a convolutional neural network approach how to design different band structures and successfully predict their topological properties purely based on the choice of unit-cell geometry. Within this context their neural network--much like the one we will discuss in this paper--constructed a tight binding Hamiltonian. We stress that their work differs significantly from our work in that we put our focus explicitly on systems without translational invariance and instead of topology focus on conductivities. Our work also develops the mathematical mappings between tight binding models and convolutional layers.
  
  Another work closely related to ours by Yu {\it et al.} \cite{Yu_2020} employs a convolutional neural network approach to make predictions about localization in disordered lattice systems and the inverse problem of predicting possible disorder configurations from localization properties. The focus of their work differs significantly from ours in that their predicted quantities do not depend on additional input parameters, such as energy, that complicate predictions and therefore necessitate a different network structure - as we will see later. Moreover, our work highlights the important insights that can be gained from the inner workings of the neural network structure. Li {\it et al.} \cite{PhysRevB.102.064205} use a neural network approach to study the conductivity in a quasi-1D wire with a small scattering region with disordered on-site energies, but do not study energy or valley-resolved conductivities, nor realistic shape deformations. Finally, a work by Torres {\it et al.} \cite{Torres2019Vnfi} used a feed-forward neural network to study valley-resolved transport in quasi-1D nanobubble superlattices. The focus of this work was limited to these superlattices, which is outside the scope of our work.
  
  The manuscript is structured as follows: In Sec. \ref{methods}, we detail the tight binding Hamiltonian and in Sec. \ref{sec:neural_network_architecture} we discuss the neural network architecture that was chosen and compare its results quantitatively to other related architectures. In Sec. \ref{networkresults}, we analyze the performance of the network in predicting total and valley-resolved conductances. In Sec.\ref{sec:graph_numeric} we discuss a graph theoretical mapping between a convolutional neural network layer and a tight binding Hamiltonian, as well as a complementary numerical result of the network. Lastly, in Sec.\ref{robust} we test the robustness of the trained neural network by evaluating its performance on deformations deviating from the Gaussian deformations that were used to train the network.

\section{Model}\label{methods}
While many of our methods are broadly applicable to questions in materials physics, we consider a specific and popular model to demonstrate our methodology. We consider transport properties of a deformed graphene flake that has dimensions of 200 by 200 lattice sites and is connected to semi-infinite leads. For simplicity units are chosen such that $ a = 1 $ (if $a = 2.46$\AA, a side length of the flake is about 50nm). Given the potential applications of nanoscale devices, we chose to investigate a system of comparable size. With no deformation, this system has a conductance quantized in units of $ \frac{2e^2}{h}$ ($e$=electron charge, $h$=Planck's constant). The introduction of random out-of-plane deformations changes this conductance profile, with a more complex relationship emerging between the deformations and the conductance as a function of energy.

Electronic transport in the graphene system is modeled using a tight-binding Hamiltonian \cite{altland_condensed_2010}, written in second-quantized form as  
\begin{equation}
     \hat{H} = \sum_{\langle i,j\rangle,\sigma} t_{ij} (\hat{a}_{i,\sigma}^{\dagger}\hat{b}_{j,\sigma} + \hat{b}_{j,\sigma}^{\dagger}\hat{a}_{i,\sigma}),
     \label{eq:graphene_ham}
 \end{equation}  
where we sum over all nearest neighbor sites $i$ and $j$ and spin projections $\sigma$. The operators $ \hat{a}_i$ and $ \hat{a}_i^{\dagger}$ are fermionic creation and annihilation operators operating on site $i$ in sublattice A, while $ \hat{b}_j$ and $ \hat{b}_j^{\dagger}$ equivalently operate on site $j$ in sublattice B. Lattice deformations locally alter the distance $d_{ij}$ between sites $i$ and $j$. This is modeled by a distance-dependent hopping parameter \cite{pereira_tight-binding_2009,settnes_patched_2015}, 
 \begin{equation}
     t_{ij} = - t_0\exp\left(-3.37 \left| \frac{d_{ij}}{a}-1\right|\right).
     \label{eq:hopping}
 \end{equation} 
 
  We choose units such that the initial hopping parameter is $t_0 = 1$. This model of transport is implemented in KWANT, a quantum transport package in Python \cite{groth_Kwant_2014}. In KWANT the system is initialized as a graphene nanoribbon with a 200x200 unit scattering region and two semi-infinite leads. 
 We use KWANT to obtain the scattering matrix $S$. The submatrix corresponding to transmission from the left lead to the right is $s=S_{LR}$, allowing computation of the total left-to-right transmission probability by the Fisher-Lee formula \cite{fisher_relation_1981},
 $ T = {\rm Tr} \left(s^{\dagger}s\right)$. The transmission probability is related to the conductance by $G = \frac{2e^2}{h} \times T$, where the factor of two emerges from spin degeneracy.
 
   By identifying the momenta of the modes corresponding to each element of $s$, it is possible to separate $s$ into submatrices corresponding to transmission and scattering between the K and K' valleys,
  \begin{equation}
     s = \begin{pmatrix} s_{KK} & s_{KK'} \\ s_{K'K} & s_{K'K'} \end{pmatrix}.
 \end{equation}
 This allows separation of the left-to-right transmission into the valley contributions,
 \begin{equation}
    T_{\alpha\beta} = {\rm Tr}\left(s_{\alpha\beta}^{\dagger}s_{\alpha\beta}\right).
 \end{equation}
 The number of conductance modes is given by $2n+1$, where $n$ is the number of occupied subbands at energy $E$ \cite{castro_neto_electronic_2009}, and we observe a 2-fold valley degeneracy and a single edge mode coming from the zig-zag edges. The transmission probabilities are normalized for each mode, so for an undeformed system the conductance is quantized and given by $G(E) =\left(2n+1 \right) \frac{2e^2}{h}$.
 
\section{Neural Network}
\label{sec:neural_network_architecture}

\subsection{Network Architecture}
The neural network developed for this investigation is a mixed input neural network with a CNN branch and is implemented in TensorFlow \cite{abadi_tensorflow_2016}. The network consists of a convolutional branch and a sequential branch to process a second round of inputs. The convolutional branch has a design that is loosely based on the AlexNet image recognition network \cite{krizhevsky_imagenet_2017}. A unique aspect of our network design is the redundant input of energy; energy is input to the sequential branch, but also included in the convolutional input, which consists of a (100,100) heightmap array and an array of the same size with every element equal to the energy. This input array with dimensions (100,100,2,1) is fed into the convolutional branch, and successive rounds of convolution, pooling, and normalization are applied.

The outputs of the two network branches are joined and analyzed in a final series of dense layers to produce final predictions for conductances. The parameters of the model are optimized using the ADAM variant of gradient backpropagation \cite{lecun_efficient_1998}. Additional specifications of the neural network architecture are found in Table \ref{tab:add_spec_netw}. 
\begin{table}[H]
\begin{tabular}{ |p{4cm}||p{4cm}|  }
 \hline
 \multicolumn{2}{|c|}{\textbf{Additional Specifications}} \\
 \hline
  \multicolumn{2}{|c|}{Architecture} \\
 \hline
Activation Function & swish function \\
Kernel Initializer & He Uniform \\
Convolutional Kernel Size & (3,3,2)\\
Average Pool Size & (2,2,1) \\
Dropout & 0.5 after dense (2048, 256)\\ Padding & Zero padding\\
Training dataset & 26,250\\
Test dataset & 8,750\\
 \hline
  \multicolumn{2}{|c|}{Training} \\
 \hline
 Optimizer & Adam \\
 Learning rate & 0.001\\
 $\epsilon$ & $10^{-7}$ \\
 Training metric & Mean Squared Error\\
 Batch size & 16\\
 Epochs & 100\\
 \hline
\end{tabular}
\caption{Neural network architecture and hyperparameters chosen for the optimal model.}
\label{tab:add_spec_netw}
\end{table}

\subsection{Optimal Design Choices}

It is also important to understand the choices that have led to our specific type of network. The optimization of this neural network required trial-and-error variations of the hyperparameters and architecture. Some such variations reinforced standard choices; for example, the optimal progression and geometry of convolutional layers is identical to that found in an image recognition network such as AlexNet.

Other common neural network design principles succeeded, too. Dropout and batch normalization layers were found to be essential to the success of this network. Batch normalization is implemented after convolutional layers, normalizing the output. The reason batch normalization works is disputed, but current theories propose that these layers may smooth the landscape of the loss function \cite{santurkar2018does} or reduce undesirable covariate shifting of neural network parameters \cite{ioffe2015batch}. Dropout layers, meanwhile, are applied after dense or fully connected layers. Dropouts randomly set some proportion of the data points to zero, effectively introducing noise. They are effective in preventing overfitting \cite{srivastava2014dropout}, where a network essentially memorizes training data and cannot successfully generalize to unseen data.

The Adam optimizer \cite{DBLP:journals/corr/KingmaB14} is chosen for its known strengths compared to other optimization algorithms, notably in reaching a compromise in speed and accuracy, and avoiding a vanishing gradient. We observed optimal performance when training with the default parameter values in Tensorflow. A less standard feature we employ is the newly developed swish function for nonlinear activation \cite{tripathi_swish_2019}. While ReLU is the more common alternative, we found swish to have better performance. The swish function is smooth and non-monotonic, which is thought to give an advantage in avoiding vanishing gradients \cite{ramachandran_searching_2017}--when the gradient of the loss function goes to zero it inhibits learning.

\subsection{Redundant Inputs} \label{ssec:redundant}
 The impact of redundancy in neural networks has been explored both in the context of the biological origins of these networks in the brain \cite{medler_training_1994}, and in direct applications in physics \cite{agliari_neural_2020}. We report here a marked improvement of network performance with the inclusion of redundant inputs. 

The highly nonlinear behavior neural networks exhibit make it challenging to understand success in network architecture beyond empirical findings. The introduction of a repeating energy array as a convolutional input does not entirely follow the intuition behind these networks, as there is no spatial variation in this constant input. The important distinction to be made is that the heightmap and energy array are input together. Consider the introduction of deformations in a lattice: This will result in local changes in potential, and the energy will directly determine the impact of these changes on the electronic transport. The stacked energy and height arrays, combined with a 3D convolutional kernel, can be imagined as a local comparison of the potential landscape with our known electronic energy. The inclusion of the small sequential branch near the end of the network can be thought of as a "reminder" of this energy input, as the exact numerical value of the energy is encoded only implicitly within the convolutional output.

We now evaluate empirically the importance of redundant inputs. We tested the network's performance given the omission of each one of two redundant input paths. Details on network training is discussed later in Sec. \ref{networkresults} of the manuscript and is omitted here for brevity. Let us briefly summarize that the final version of the network seen in Fig. \ref{fig:net} has a mean absolute error (MAE) of $ 0.61 \frac{2e^2}{h}$ for predicted total conductance (see Sec. \ref{sec:total_conduct} for more details), with predictions made on new, {\it i.e.} untrained data. We found that omitting the energy array as a convolutional input (marked A in Fig. \ref{fig:net}) causes the MAE to increase to $ 0.87 \frac{2e^2}{h}$, while omission of the small second branch input (marked B in Fig. \ref{fig:net}) results in a MAE of $ 0.98 \frac{2e^2}{h}$. Comparison to the performance of the complete network shows just how important the network redundancy is - after all, the added redundancy in network architecture leads to a decrease of more than 30\% for the MAE. We share this finding with the intent of providing more design intuition for physics applications of NNs.

\begin{figure*}
    \centering
    \includegraphics[width=\textwidth]{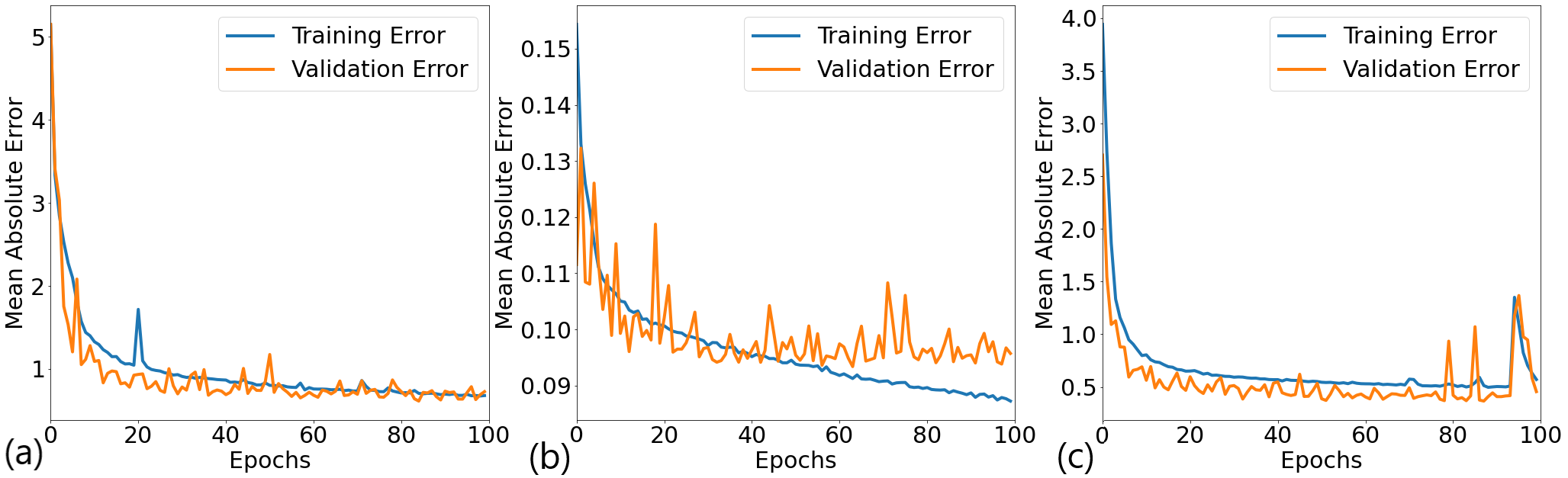}
    \caption{Network error on training data and validation data for (a), the total transmission; (b) the off diagonal components of transmission; and (c) the diagonal components of transmission.}
    \label{fig:met}
\end{figure*}

\section{Training and Results}\label{networkresults}

For the training data of our neural network we consider random deformations of the lattice, which are modeled using 2D Gaussian bumps that are randomly placed. For concreteness in our case the number of gaussians $N$ is chosen uniformly from $N \in [1,10]$. For each Gaussian, Eq.\eqref{eq:Guassian}, parameters $(A,\sigma_x,\sigma_y,x_c,y_c)$ are chosen uniformly from the ranges in Table~\ref{tab:gauss_param} below, including the approximate values in nanometers(nm) for graphene, $a = 2.46$\AA.
\begin{table}[H]
\begin{tabular}{ |p{4cm}||p{4cm}|  }
 \hline
 \multicolumn{2}{|c|}{\textbf{Gaussian Parameter Bounds [a]/[nm]}} \\
 \hline
A & (0,10) / (0,2.5)\\
$\sigma_x, \sigma_y$ & (5,20) / (1.5,5) \\
$x_c, y_c $ & (-60,60) / (-15,15)\\
 \hline
\end{tabular}
\caption{The numerical bounds for the amplitude, standard deviation, and center of each Gaussian bubble. Values are provided in both lattice units and nanometers.}
\label{tab:gauss_param}
\end{table}
These values are chosen in accordance with previous theoretical literature \cite{wurm_symmetries_2012,settnes_graphene_2016}. The random superposition of these deformations produces a net deformation comparable to small graphene nanobubbles observed experimentally, less than 50nm in radius \cite{ghorbanfekr-kalashami_dependence_2017,aslyamov_model_2020,kim_initial_2021}. 

More precisely, this means that we model the height of a point $(x,y)$ on the graphene sample by,
 \begin{equation}
    z(x,y) = \sum_{n=1}^N A_n \exp \left( -\frac{(x-x_{cn})^2}{2\sigma_{xn}^2} - \frac{(y-y_{cn})^2}{2\sigma_{yn}^2} \right).
    \label{eq:Guassian}
 \end{equation}
To generate the height maps that are used as inputs of the neural network this expression is evaluated over a 100x100 grid spanning the scattering region. Eq.(\ref{eq:Guassian}) is also used by KWANT in conjunction with Eq.(\ref{eq:hopping}) to construct the Hamiltonian and obtain conductance values for each sample at a random energy. We focus on energies in the first 50 subbands, corresponding to the first 99 conductance modes. These conductance values are the target for the network, which continually evaluates its performance and uses gradient backpropagation to improve the model parameters. A dataset of 35,000 samples was generated in KWANT, with 75\% used for supervised learning and 25\% used to validate the network accuracy. Networks were trained to learn the total left-right transmission $T$ as well as the valley-resolved transmission components $T_{\alpha\beta}$.

\subsection{Details of network training}

Plotting the ``learning curve" of the networks, Fig. \ref{fig:met}, we see the error of the model on both the training data and validation data evaluated over every epoch, or cycle, through the data. For all of these plots, the validation error appears to converge to a fixed value, a broad indicator of successful training . Tall, narrow spikes in the error may be observed, but this does not appear to cause any problems, as the error quickly returns to the convergent value. There is an expected trade off with model stability and model generalization error when modifying batch size: large batches result in a smooth, stable curve and validation error that will tend to converge higher, while small batch sizes give better generalization with smaller validation errors, but are more volatile \cite{masters_revisiting_2018,radiuk_impact_2017}. Training neural networks is a complex task, and as such this topic is much broader than what we discuss here.

\subsection{Total Conductance}
\label{sec:total_conduct}
\begin{figure}[t]
    \centering
    \includegraphics[width=8cm]{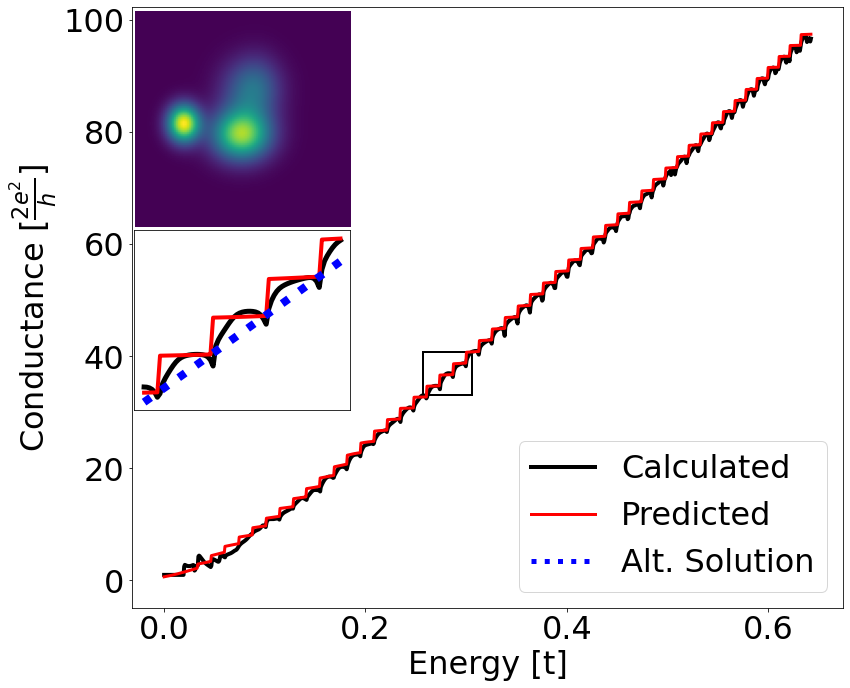}
    \caption{ The calculated and predicted conductance of a single deformed sample over a range of energies. The upper inset panel shows the relative height of the deformed sample, and the panel below provides a closer look at the predictions in the small boxed region. The blue dotted line shows the solution the network provides when trained with the non-quantized model. }
    \label{fig:totcond}
\end{figure}

After 100 epochs of training, the validation error (the network's error on new data not encountered in training) reaches 0.61 $\frac{2e^2}{h}$ mean absolute error (MAE), in a dataset with mean conductance of roughly 41$\times \left(\frac{2e^2}{h}\right)$. This is a 4.3\% average relative difference, based on the formula:
 \begin{equation}
    \frac{\left | y_p - y_c \right |}{\left | y_p \right |}\times 100,
    \label{eq:error}
 \end{equation}
where $y_c$ is the calculated value and $y_p$ is the predicted value. To further illustrate the network's ability, the model is evaluated on an individual sample at 1000 linearly spaced energy values. Representative results are shown in Fig. \ref{fig:totcond} and demonstrate the network's success in learning the dependence of conductance on both deformations and energy. An alternate model which does not take mode numbers as inputs is seen in the inset. The appearance of discrete steps in conductance from the inclusion of mode numbers is a good illustration of the fine tuning of model function that is possible with variation in network structure. These models have comparable errors, so we choose the discrete model, taking mode numbers as an input, as the primary model to perform further analysis.

\subsection{Valley-Resolved Conductance}

We next train a neural network on all components of $T_{\alpha\beta}$. The valley resolved transmission brings additional challenges to the model; computation of the off-diagonal components $T_{KK^\prime}$ and $T_{K^\prime K}$ frequently gives results with large fluctuations (x2 or more) over very short energy ranges. This can be attributed to the disorder introduced by the deformations and is especially prominent at higher energies \cite{wurm_symmetries_2012}. Additionally, these components may be near-zero. Both are potential problems for the gradient backpropagation algorithm. To address this, two separate models are trained, one for the valley transmissions $T_{KK}$ and $T_{K^\prime K^\prime}$, and one for the inter-valley scattering components $T_{K^\prime K}$ and $T_{KK^\prime}$. To get precise predictions even for the small off-diagonal scattering components $T_{K^\prime K}$ and $T_{KK^\prime}$ we scale them by a factor of $10^6$ before training (predictions are divided by the same factor for comparison). The error in this approach is found to reduce significantly from the unscaled case, decreasing from 0.15 to 0.095 $\frac{2e^2}{h}$. This result is a consequence of the canonical issue of the ``vanishing gradient," which we otherwise largely avoided by use of the Swish activation function \cite{ramachandran_searching_2017}. 
\begin{figure}[b]
    \centering
   \includegraphics[width=8cm]{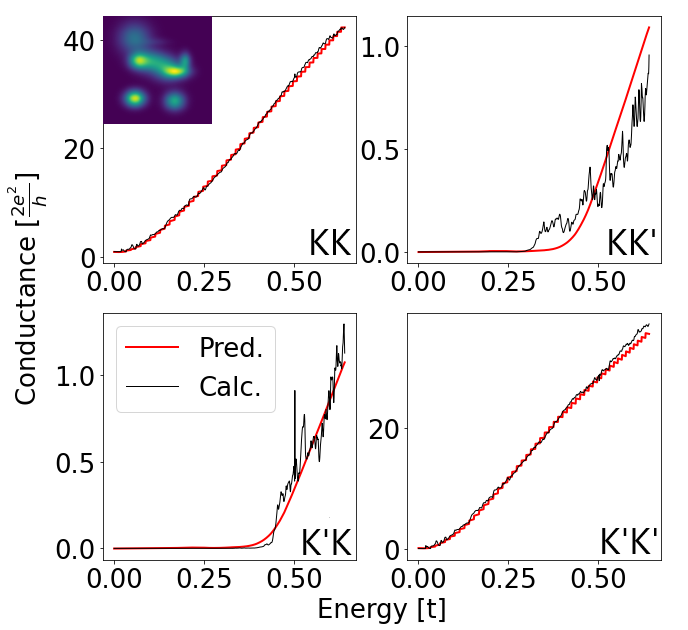}
    \caption{The calculated (red) and predicted (black) valley-resolved conductance of a single deformed sample over a range of energies. The inset shows the height map of the sample in question.}
    \label{fig:rescond}
\end{figure}

Our method allows for the successful prediction of all four components of the transmission. The average value of each component and the mean absolute error in the prediction for each component is,
\[ \langle T_{\alpha\beta}\rangle = \begin{pmatrix} 19.79 & 0.28 \\ 0.28  & 20.66  \end{pmatrix}, \; 
M.A.E. = \begin{pmatrix} 0.37 & 0.094  \\ 0.095  & 0.36  \end{pmatrix}, \]
expressed in units of 2$e^2/h$, where $\langle T_{\alpha\beta}\rangle$ is the average transmission matrix over all samples and energies, and $M.A.E.(T_{\alpha\beta})$ is the average prediction error over all samples and energies. This metric of error is just one measure of model success, but in combination with plotted predictions it may provide some intuition as to the performance of the model. The mean error corresponding to inter-valley scattering is quite small numerically, which can be attributed to the often near-zero value of these components. 

To better understand model performance, another prediction for a single sample is included in Fig. \ref{fig:rescond}. The $KK$ and $K^\prime K^\prime$ components of transmission were predicted successfully comparing the magnitude of the error with the total conductance model. For the off-diagonal components representing inter-valley scattering, the model's approximation succeeded with an average error of 0.095 $\frac{2e^2}{h}$, and this success can be seen qualitatively in Fig. \ref{fig:rescond}. This example  demonstrates that this model can not only predict the magnitude and trend of inter-valley scattering, but can also predict whether it occurs at all, and at what energy these effects become significant.

\section{Graph theoretical interpretation}
\label{sec:graph_numeric}
In this section we use a graph theoretical perspective to illuminate an interesting mathematical connection between the tight-binding Hamiltonian and a layer of a CNN, finding that the honeycomb lattice graph is isomorphic to a subgraph of the convolutional layer graph. Following this, we provide numerical evidence that is complementary to this finding; in Appendix \ref{sec:A} we demonstrate that the graph structure and numerical outputs together can be used to re-construct a tight binding Hamiltonian. This is a valuable theoretical contribution to the existing design principles of neural networks for applications in physics. For a detailed definition of graph terminology and notation, see Appendix \ref{sec:A}.

\subsection{Honeycomb lattice as a subgraph of the convolutional layer}

Consider a single convolutional layer in a neural network. In a 2D convolution, a kernel is passed over the values of an array, transforming some $N \times N$ input $\mathbf{X}$ to a $N \times N$ output $\mathbf{Y}$. We consider here a 3x3 kernel, as implemented in our neural network. The relationship between the input array $\mathbf{X}$ and the output array $\mathbf{Y}$ is as follows. At the index $(m,n)$, the output value $\mathbf{Y}_{m,n}$ is given by:

  \begin{equation}
\mathbf{Y}_{m,n} = \sigma\left(\sum_{i,j=-1}^{1} w_{ij}\mathbf{X}_{m+i,n+j}+\beta_{ij}\right),
\label{eq:conv}
 \end{equation}

We can see that, in the language of tight-binding, the output $\mathbf{Y}_{m,n}$ is dependent on $\mathbf{X}_{m,n}$ and its first and second nearest neighbors on the square input grid. This extends naturally to a formulation in graph theory.
\begin{figure}[t]
    \centering
    \includegraphics[width=6cm]{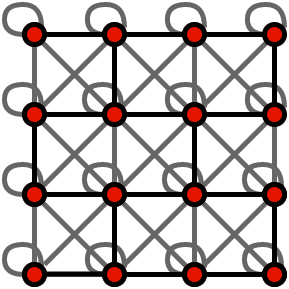}
    \caption{Honeycomb lattice (black edges) as a subgraph of the convolutional layer graph (all edges)}
    \label{fig:brickwalla}
\end{figure}

Define a graph $G$ with the adjacency matrix $\mathbf{A}(G)$, given:

\begin{equation}
\mathbf{A}_{ij}(G) =
\begin{cases}
 
1 & \hspace{5mm} j \in e(i) \\
0 & \hspace{5mm} j \notin e(i) \\

\end{cases},
\label{eq:adj}
 \end{equation}
 where $e(i)$ is the set of nodes with which node $i$ shares an edge. Based on Eq.(\ref{eq:conv}), in the formalism introduced by You {\it et al.} \cite{you_graph_2020}, the sets $e(i)$ defining the convolutional graph $G_C$ are given by the set of points enclosed by a 3x3 kernel centered at a node $(m,n)$:
 
\begin{equation}
k_{3x3}(m,n) = \{ u \in (m-1,m+1), v \in (n-1,n+1)\}.
\end{equation}

Meanwhile, the sets $e(i)$ defining the graph of a honeycomb lattice $G_g$ are given by the sets of first nearest neighbors,

\begin{equation}
N_{1}(m,n) = \{(m-1,n), (m+1,n), (m,n\pm1\}.
\end{equation}

By the definitions of these edge sets, we see that:

\begin{equation}
N_{1}(m,n) \subseteq k_{3x3}(m,n).
\end{equation}

And thus, when both graphs are defined on a $N\times N$ grid of points, the honeycomb graph $G_g$ is a subgraph of the convolutional layer graph $G_C$. This is depicted in Fig. \ref{fig:brickwalla}. This demonstrates simply that the graph structure of a first nearest-neighbors tight binding Hamiltonian on a honeycomb lattice is included within the graph structure representing the action of our first convolutional layer.

\subsection{Linear Map}
Given this graph structural parallel, it is instructive to ask whether the numerical inner workings of the neural network also show some parallel to our tight-binding model. 

An interesting work by Sun {\it et al.} \cite{sun_deep_2018} found that in a CNN trained to predict Chern numbers from Hamiltonians, the Berry curvature in momentum space was approximately recreated, as an intermediate step. This was taken to indicate the success of the CNN in recreating the mathematical steps between input and output.

\begin{figure}[t]
    \centering
    \includegraphics[width=8cm]{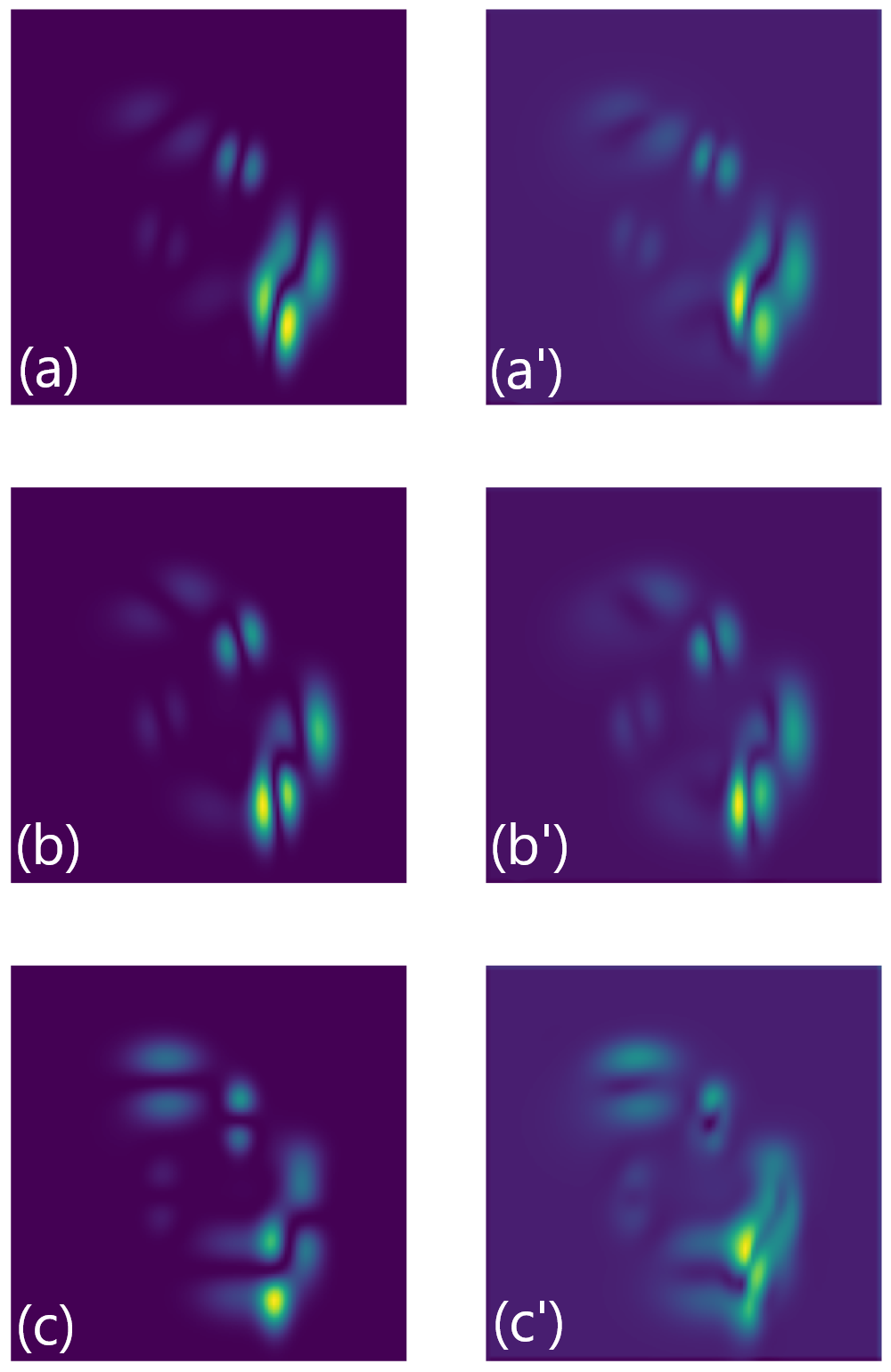}
    \caption{The exact nearest-neighbor hoppings as used in KWANT [for the various directions $a=(\sqrt{3}/2,-0.5)$, $b=(-\sqrt{3}/2,-0.5)$, $c=(0,1)$], and the result extracted via the linear mapping linear mapping from the network intermediate output in Fig. \ref{fig:net} (a',b',c')}
    \label{fig:tb}
\end{figure}

We similarly studied the intermediate outputs of each convolutional layer - after activation and batch normalization. We find that for any deformed sample, the set of feature maps $\mathbf{F}$ output from the first CNN layer approximately satisfies a linear map $f: \mathbf{F} \rightarrow \mathbf{h}$  to the calculated hopping amplitudes in each direction $\mathbf{h}$, such that:
 \begin{equation}
\mathbf{h} = \mathbf{A}\mathbf{F},
    \label{eq:map}
 \end{equation}
where $\mathbf{F}$ is a 16-component vector of outputs at some array index, $\mathbf{A}$ is a 3x16 transformation matrix, and $\mathbf{h}$ are the 3 hopping amplitudes in the nearest-neighbor directions, calulated with Eq.(\ref{eq:hopping}). This is illustrated in Fig. \ref{fig:tb}. We focus on only one of the two arrays output by this network layer. The inclusion of the other 100x100 component of this output, corresponding to the energy array input, does not change appreciably the result of this linear map, so we omit it for simplicity.

We find this linear map can recreate the calculated hoppings at an average of 1\% error. This mapping is the simplest way for 16 output values to encode 3 nearest-neighbor hoppings. While any arbitrarily complex function could extract hopping values from these outputs, their appearance as a simple linear combination shows that this information is encoded in this output, implying the network truly learns how to construct hopping parameters from a deformation image.

We have shown that the honeycomb lattice graph $G_g$ is a subgraph of the convolutional graph $G_C$. Consider now the linear map $f$ we have introduced here. In Appendix \ref{sec:A}, we show it is possible to use this linear map, in conjuction with the graph relation discussed above, to form a complete representation of the tight binding model, from the structure and output of the first convolutional layer of our neural network. When we use this linear map to assign weights to the appropriate entries in the adjacency matrix, it is possible to fully construct the matrix of the tight-binding Hamiltonian.

\subsection{Discussion of Extensions and Limitations}

In the general consideration of physical problems with an inherent graph structure using neural network methods, the question remains: does the graph of a convolutional layer and the local ``interactions" it depicts sufficiently represent the important characteristics of, in our example, electronic transport on an arbitrary 2D lattice? Or, does the problem necessitate an exactly or nearly isomorphic graph structure, as is possible in the more recently developed techniques of graph neural networks \cite{DBLP:conf/iclr/KipfW17}? 

Our network succeeded in predicting conductance values without this completely isomorphic graph structure. In fact, our tight binding model was defined on a far greater number of atomic sites than there were input data points to the convolutional layer: the success of this coarse-grained approach suggests that is is not necessary for a neural network to have a graph structure exactly isomorphic to the objective problem.

It is valuable to consider more complex tight binding models. While we leave the further application of these neural network methods to future work, let us discuss the implications of our graph theoretical insights when applied to different tight binding models.

In theory, if the convolutional kernel is allowed to be of arbitrary size, there will be a subgraph isomorphic to any tight binding model; this is apparent in the limit in which each node is connected to all others. In Appendix \ref{sec:A}, we show that the first nearest-neighbors kagome lattice is isomorphic to a subgraph of the convolutional graph. We additionally investigate in Appendix \ref{sec:A} the extension of the graph based interpretive scheme to second and third nearest neighbor graphs on a honeycomb lattice, and we come across a complication; in Fig. \ref{fig:brickwalla} we can see that comparing different sites, a nearest neighbor vector alternates between the site directly above and below. The site in the other direction represents a third nearest neighbor. This introduces an ambiguity which may be difficult for a neural network to properly resolve.

Convolutional neural networks are versatile enough that they could succeed to some degree even with these more complex cases; However, in searching for an optimally designed network, it is worth further investigating techniques catering to the exact graph structure of the problem at hand, such as graph neural networks.

While we have probed the graph structure and numerical output of the first layer of our neural network, this makes up a small portion of the network's total numerical operations. The successive pooling and convolution operations present an interesting and complex mathematical structure, but this complexity renders the network as a whole difficult to interpret rigorously. Within the framework we have developed, if the first convolutional layer represents hoppings between adjacent lattice sites, an average pooling operation and second convolutional layer then represent hoppings between 2x2 sub blocks of lattice sites.

Further development of this structure-based understanding of neural network design may provide interesting insights into this otherwise opaque topic. This is especially true in the application of neural networks within the field of condensed matter, given the ubiquitous appearance of graph structure in the study of physics on a crystal lattice.

\section{Robustness of the trained network}
\label{robust}
\begin{figure}
     \centering
     \begin{subfigure}[b]{0.45\textwidth}
         \centering
         \includegraphics[width=\textwidth]{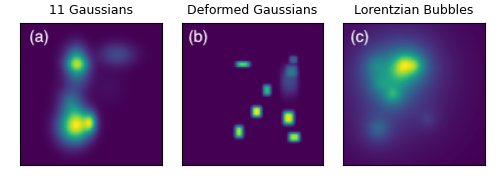}
     \end{subfigure}
     \hfill
     \begin{subfigure}[b]{0.45\textwidth}
         \centering
         \includegraphics[width=\textwidth]{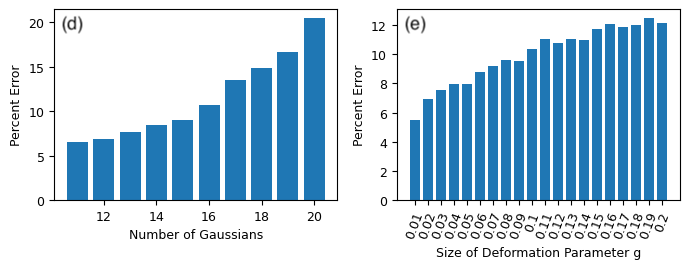}
     \end{subfigure}
     \hfill
        \caption{(a-c) Examples of different samples given to the network. (d, e) Total conductance error increases with the number of Gaussians, and the size of the deformation parameter $g$. Samples were sorted into 20 bins, each spanning a range of 0.01 in the average deformation parameter $g$. Each bin contains \(>\)6000 samples. The percent error for each bin is calculated according to Eq. \eqref{eq:rel_diff_max}.}
        \label{fig:Robustness}
\end{figure}

Next, we want to determine the quality of predictions for samples that are fundamentally different from those on which the network was trained. First, we use the same formula and parameters as the training data to generate Gaussian deformations, but we add more deformations to increase the complexity of the overall sample. The error in the networks predictions--calculated according to Eq.\eqref{eq:error}--increases as more Gaussians are added, as shown in Fig. \ref{fig:Robustness}. For example, in samples with 15 Gaussians the average error in the total conductance increases to 9\%, compared to 6.5\% for samples with 11 Gaussians. This indicates that the network becomes less accurate when faced with data that differs too drastically from training data.

Next, we want to see how robust the network is against changes to the shape of the deformations. Therefore, we test the network on deformed Gaussian bubbles of the form,
 \begin{equation}
    z(x,y) = e^{\frac{-(x-x_c)^2-c(x-x_c)^4}{2\sigma_x^2}+\frac{-(y-y_c)^2-d(y-y_c)^4}{2\sigma_y^2}}.
 \end{equation}

 To obtain better quantitative insights we introduced deformation parameters $c$ and $d$ that are randomly selected from the range $[0,0.2]$. To obtain a single metric for the overall deformation of the sample, the deformation parameters are averaged according to Eq. \eqref{eq:av_def_param}, 
 \begin{equation}
    g = \frac{1}{2N}\sum_{n=1}^N c_n+d_n.
    \label{eq:av_def_param}
 \end{equation}
 
 Here, instead of a percent error we compute the relative difference according to Eq.\eqref{eq:rel_diff_max},
 \begin{equation}
      \frac{\left | y_p - y_c \right |}{\mathrm{max}(\left | y_p \right |,\left | y_c \right |)}\times 100.
      \label{eq:rel_diff_max}
 \end{equation}
 This is done to avoid misleading results: very small calculated or predicted values that appear in a tiny fraction of samples can result in excessively large percent errors despite both results being sufficiently close to zero.  See Fig. \ref{fig:Robustness}.


Finally, we test the network on bubbles that are not Gaussians but Lorentzian bubbles of the form
 \begin{equation}
    z(x,y) = \frac{1}{\pi}\frac{\Gamma_x\Gamma_y }{(x-x_c)^2+(y-y_c)^2+\Gamma_x^2+\Gamma_y^2},
    \label{eq:lorentz}
 \end{equation}
where \(\Gamma\) is the half width at half maximum of the distribution. \(\Gamma\) is randomly chosen from the range [4,16] so that the Lorentzian bubbles are roughly the same size as the Gaussian bubbles. When tested on approximately 8800 samples the network performed very well, returning an average error (calculated according to Eq.\eqref{eq:error}) of only 2\% in the total conductance, despite the fact that the network was trained on Gaussian deformations, not Lorentzian ones. 

We can conclude from this section that the network is sufficiently robust against changes that it can be used in applications to real world data, such as one could obtain from an STM where deformations might not be perfectly Gaussian. The robustness of the trained network can be explained to some extent by the previous section, in which we showed that the network forms a tight binding Hamiltonian as an intermediate step in its predictions. These results indicate that the network has actually learned about the underlying physics, rather than just learning the geometry of Gaussian deformations.

\section{Conclusion}
This work provides a proof-of-concept that neural networks--convolutional neural networks in particular--can serve as a tool to expedite determination of the physical properties of a material. We developed a neural network capable of identifying the conductance of deformed graphene nanoribbons to within 5\% when given the heightmap, energy, and number of conductance modes at that energy. We further find that despite the fact that the network was trained with a fixed range of Gaussian deformations, predictions still remain accurate when the quantity and type of deformation are varied, indicating a robust model. We found that once trained, our model can predict conductance with the computational time reduced by a factor of $\mathcal{O}(10^{4})$ compared to an exact calculation, and it requires $\mathcal{O}(N)$ parameters in the description of a tight-binding system with $N$ sites, as compared to the $N^2$ parameters required to construct the dense Hamiltonian matrix. The desired behavior of the model can be additionally tuned with the choice of training inputs, where inclusion of the mode numbers give a semi-quantized prediction, and omission gives a smooth prediction. We also demonstrated the model's ability to learn and predict the valley-resolved components of conductance with little modification to methodology.

Additionally, we gained insight into the model's design and function by studying the internal outputs and graph structure of the neural network. It was found that the graph of the tight binding model determined from the CNN is a subgraph of the relational graph describing a convolutional layer. We conclude that this sort of fundamental structural equivalence is an important factor in the success and efficiency of this model. This should be considered in the application and design of convolutional and other deep learning networks to further research in physics.

It should be noted that different problems often require differently structured networks to obtain the best solution, but the model we developed is optimal for the problem of deformation-dependent conductance. We have shown that many of the techniques originally developed for image recognition networks can be readily adapted for this class of problems. The method described here could be tested by changing the desired output observable, or by applying this technique to different materials, such as 3D lattices or nanostructured materials. 

There are also some limitations of this method that are worth noting. We find that including the number of conductance modes as an input results in a semi-quantized output. This is preferable when considering minimal deformations, wherein the conductance retains some quantization as in the zero-deformation limit. More severely deformed samples do not exhibit these steps in conductance, and as such the model trained without conductance modes as an input, which outputs a smooth conductance prediction, is superior. An improved model would successfully differentiate between the behavior of these cases. This may be possible by training separate networks for the small vs. large deformation cases, or otherwise augmenting the training data to emphasize this difference.

Our work has demonstrated that neural network methods can be applied as an accurate approximation method to expedite the calculation of physical properties of materials. Furthermore, we have illustrated methods in neural network construction that may be especially beneficial for applications in physics, such as redundant input data. We also provide new insight into the mathematical relevance of convolutional networks to graph-based problems like the tight-binding Hamiltonian. The data and code supporting the results of this paper are available from the corresponding author, J.N., upon reasonable request.

\section{Acknowledgements}
 A. A. gratefully acknowledges the support of KFUPM through the grant SR191021. M.V. gratefully acknowledges the support provided by the Deanship of Research Oversight and Coordination (DROC) at King Fahd University of Petroleum \& Minerals (KFUPM) for funding part of this work through project No. SR211001. G.A.F. gratefully acknowledges funding from the National Science Foundation through the Center for Dynamics and Control of Materials:an NSF MRSEC under Cooperative Agreement No. DMR-1720595, with additional support from NSF DMR-1949701 and NSF DMR-2114825.  This work was performed in part at the Aspen Center for Physics, which is supported by National Science Foundation grant PHY-1607611. This work was also completed in part using the Discovery cluster, supported by Northeastern University’s Research Computing team.

\appendix

\section{Details of Graph Theoretical Analysis}
\label{sec:A}
\subsection{Notation and Terminology}

\begin{figure}[H]
    \centering
    \includegraphics[width=6cm]{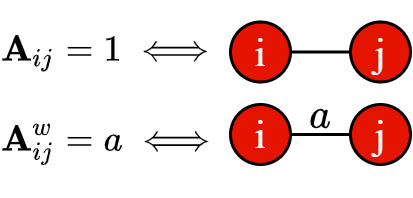}
    \caption{Adjacency matrix definition.}
    \label{fig:adj}
\end{figure}

Here we briefly define the important concepts and notation we use in our analysis. A graph is a finite set of nodes connected by a finite set of edges, where each edge connects a pair of nodes \cite{rahman_basic_2017}. In this analysis, we work with non-directed graphs, so each edge is defined simply by an unordered pair of nodes. In a tight-binding framework, nodes are equivalent to atomic sites, and edges to corresponding hopping parameters. The primary tools we will use to represent these graphs are the unweighted and weighted adjacency matrices $\mathbf{A},\mathbf{A}^w$. In $\mathbf{A}$, $\mathbf{A}_{ij}=1$ if and only if nodes i and j share an edge. In $\mathbf{A}^w$, $\mathbf{A}^w_{ij}=a$ when nodes i and j share an edge with weight $a$. These definitions are illustrated in Fig. \ref{fig:adj}. 

 In each matrix, the element $ij$ is valued at zero when no edge exists between nodes $i$ and $j$. If two graphs $G, G'$ have the same adjacency matrix $\mathbf{A}$, but not necessarily the same edge weights, we call the graphs $G$ and $G'$ structurally isomorphic, because we know the nodes and edges are identical. If we additionally know that $G$ and $G'$ have the same weighted adjacency matrix $\textbf{A}^w$, we will call these graphs completely isomorphic, because we know the nodes, edges, and weights are identical. 
 
\subsection{Convolutional Layer Representation of Tight Binding Hamiltonian}

The tight binding matrix is, as a symmetric matrix, always equivalent to the weighted adjacency matrix $\mathbf{A}^w$ of some undirected graph with edge weights, $G^w_H$. To prove a graph isomorphism extending to a honeycomb lattice graph $G_g$, we introduce edge weights to the convolutional graph, $G_C \rightarrow G^w_C$. For each node $i$ in $G_C$, there are $k$ corresponding outputs in the set of feature maps $\mathbf{F}$. Similarly, at node $i$ in $G_H$, there are up to $N$ $nth$ nearest-neighbor hoppings in the set $\mathbf{h}$, described equivalently by the nonzero values in row $i$ of the tight-binding matrix $\mathbf{H}$. When $N \leq k$, there are neural network outputs such that
\begin{equation}
    f: \mathbf{F} \rightarrow \mathbf{h}
 \end{equation}
 is a linear mapping from $\mathbb{R}^k \rightarrow \mathbb{R}^N$, which we choose to assign the edge weights of $G^w_C$. 
 
 Beyond the theoretical existence of this linear map, we discussed in Sec. \ref{sec:graph_numeric} that there is indeed such a mapping, accurate to about 1\% error, which produces the calculated nearest-neighbors hoppings from the feature map outputs with a linear map $\mathbf{h} = \mathbf{AF}$. We allow the assignment of 0 to edges, and take this to delete the edge. This enables a reduction to any subgraph of $G_C$, so we turn our attention the the graphene tight-binding Hamiltonian, represented by the weighted graph $G^w_g$, where $G^w_g$ is constructed with the tight-binding matrix as a weighted adjacency matrix $\mathbf{A}^w(G^w_g)$, such that the edge between nodes $i$ and $j$ has a weight equal to element $ij$ of the tight binding matrix:
\begin{equation}
\mathbf{A}^w(G^w_g): \mathbf{A}^w_{ij} = \mathbf{H}_{ij}.
\end{equation}

When the conditions are met for the linear map, such that the convolutional layer's output maps linearly to the set of nearest-neighbor hoppings at each point, we assign the convolutional graph edge weights accordingly,
\begin{equation}
\mathbf{A}^w(G^w_C): \mathbf{A}^w_{ij} = \begin{cases}
 
f(\mathbf{F}_i)_n & \hspace{5mm} j \in N_g(i) \\
0 & \hspace{5mm} j \notin N_g(i) \\

\end{cases},
\end{equation}
where $f(\mathbf{F}_i)_n$ is the $n^{th}$ of $N$ possible second nearest-neighbor hoppings. By the nature of the map, these are only non-zero when the $n$-th node $j$ is included within the set of brick wall first nearest neighbors at node $i$, denoted $N_g(i)$. For any $G^w_g$ there exists a convolutional layer output $\mathbf{F}$ and a linear mapping $f$ such that the assignment of edge weights of $G^w_C$ by $f$ makes $G^w_C$ completely isomorphic to $G^w_g$. 

We have demonstrated that the graph structure and numerical outputs of this neural network layer can be used to construct a complete representation of the tight-binding matrix.

\subsection{More Complicated Lattices}

\begin{figure}[h!]
    \centering
    \includegraphics[width=4cm]{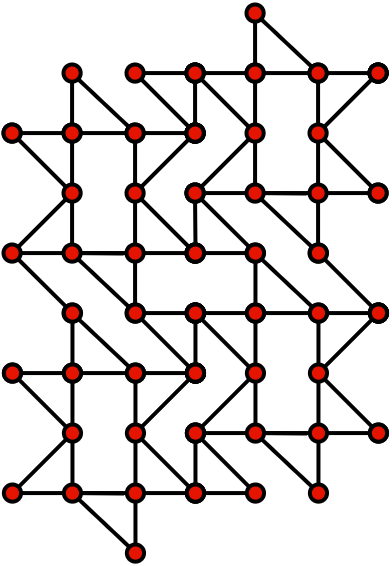}
    \caption{Kagome lattice shown as a subgraph of the convolutional graph (not shown)}
    \label{fig:Kagome}
\end{figure}

Consider a first nearest neighbors tight-binding Hamiltonian on a Kagome lattice. Mapping this to a square lattice is more challenging, but there are multiple periodic subgraphs of $G_C$  that are isomorphic to our Kagome lattice. One such subgraphs is depicted in Fig. \ref{fig:Kagome}. 

 \begin{figure}[t!]
    \centering
    \includegraphics[width=4cm]{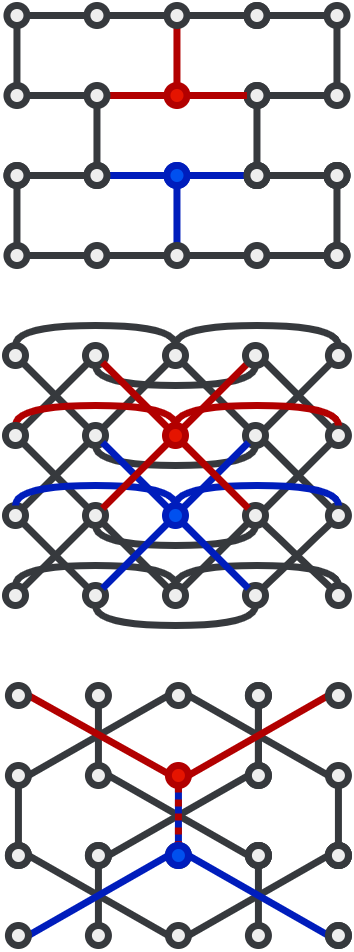}
    \caption{First (top) second (middle) and third (bottom) nearest neighbors of two points on the honeycomb lattice.}
    \label{fig:123onn}
\end{figure}

In fact, it is always possible to map a first nearest neighbor tight binding Hamiltonian on an arbitrary 2D lattice onto a subgraph of a convolutional graph with an appropriately sized kernel. This is evident in the fully-connected limit; for a $(2N+1) \times (2N+1)$ kernel on an $N \times N$ grid of points, the convolutional graph is complete, with every pair of vertices connected by an edge.

Typically, however, this upper limit is not necessary: For both a honeycomb and Kagome lattice, we find isomorphic subgraphs of the convolutional graph created by a 3x3 kernel. 

We also consider the graph mappings of second and third order nearest neighbors on the honeycomb lattice, Fig. \ref{fig:123onn} Some nearest neighbor points in the square lattice $(i,j)$ and $(i+1,j)$ alternate between first and third nearest neighbors in the honeycomb lattice mapping, while second order neighbors are symmetric between sites. The lack of translational invariance in these hoppings between the sites of the square lattice represents a breakdown of the exactness of the mapping we consider; to represent exactly these graph structures would require graph neural network techniques. However, our results suggest that this exact representation may not be necessary.

\bibliography{references}
\end{document}